\documentclass[twocolumn,showpacs,preprintnumbers,amsmath,amssymb]{revtex4}

\usepackage[dvips]{graphicx}

\usepackage{dcolumn}
\usepackage{bm}
\usepackage{amsmath}
\usepackage{amssymb}
\usepackage{color}
\usepackage{units}

\usepackage[french]{babel}
\usepackage[latin1]{inputenc}

\long\def\symbolfootnote[#1]#2{\begingroup \def\thefootnote{\fnsymbol{footnote}}\footnote[#1]{#2}\endgroup}

\begin{document}
\preprint{APS/}

\title{Analytical study of electrostatic ion beam traps}

\author{Alexandre Vallette}\email{alexandre.vallette@spectro.jussieu.fr}
\author{P. Indelicato} \email{paul.indelicato@lkb.ens.fr}

\affiliation{%
Laboratoire Kastler Brossel, École Normale Supérieure,
CNRS, Université Piere et Marie Curie -- Paris 6, Case 74; 4, place
Jussieu, 75252 Paris CEDEX 05, France}%
%

\date{Automatic Time-stamp: <Sunday, December 13, 2009, 23:10:46 >}

\begin{abstract}
The use of electrostatic ion beam traps require to set many potentials on the electrodes (ten in our case), making the tuning much more difficult than with quadrupole traps. In order to obtain the best trapping conditions, an analytical formula giving the electrostatic potential inside the trap is required.
In this paper, we present a general method to calculate the analytical expression of the electrostatic potential in any axisymmetric
set of electrodes. We use conformal mapping to simplify the geometry of the boundary.
The calculation is then performed in a space of simple geometry.
We show that this method, providing good accuracy, allows to obtain the potential on the axis as an analytic function of the potentials applied to the electrodes, 
thus leading to fast, accurate and efficient calculations. We conclude by presenting stability maps depending on the potentials that enabled us to find the good trapping conditions for $O^{4+}$ at much higher energies than what has been achieved until now.
\end{abstract}

\pacs{Valid PACS appear here}
                              
\maketitle

\section{Introduction\protect}
\label{sec:intro}

\begin{figure}
\includegraphics[width=0.3\textwidth,trim = 8cm 4cm 20cm 4cm,clip]{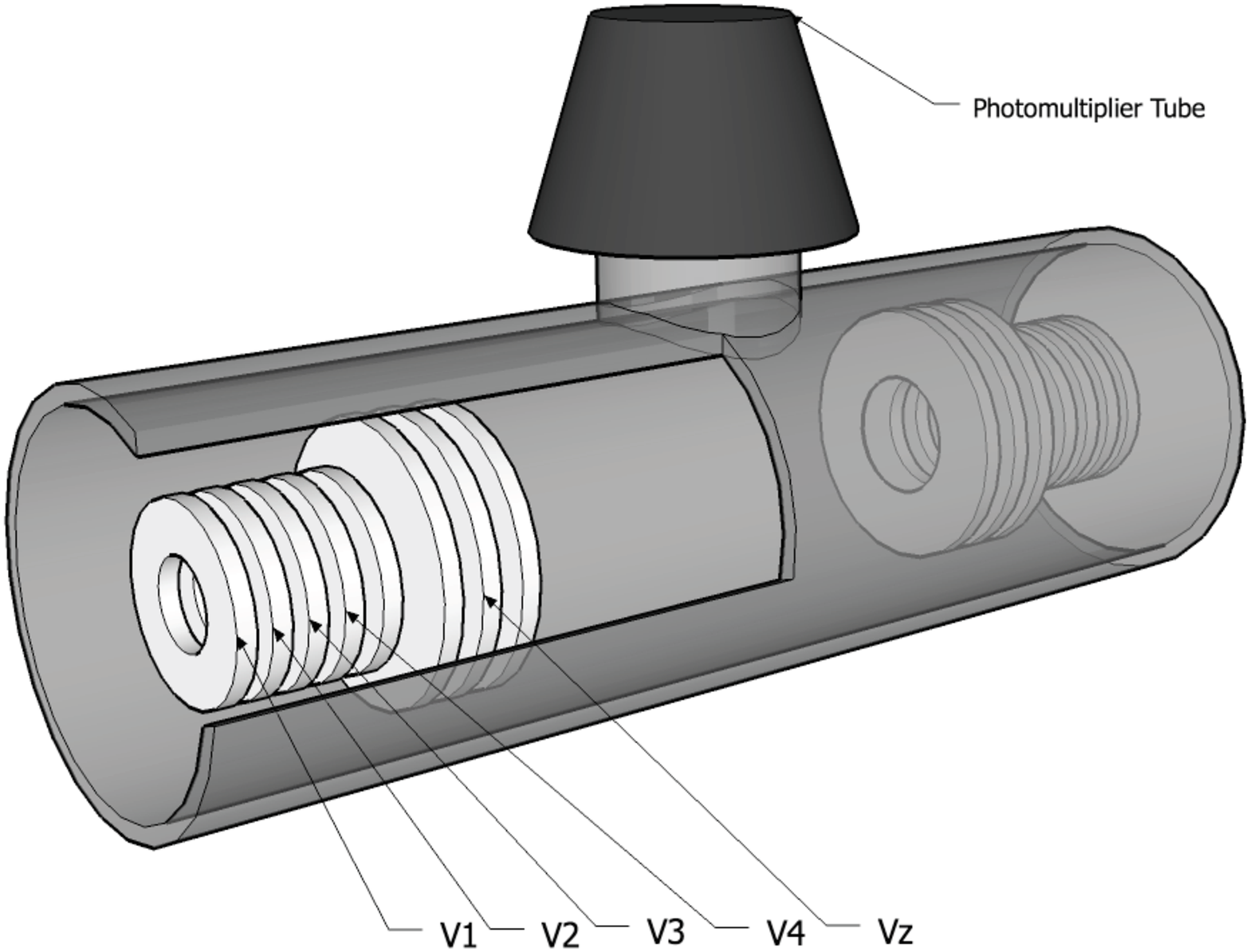}
\caption{\label{trap3d} (Color online) Overview of the EIBT. Five potentials are applied to the electrodes, the other are grounded. The injection of the bunch of ions is performed when all the electrodes on one side of the trap are grounded. The potentials are raised before the bunch has time to come back. This trap can be used to make metastable state lifetime measurements, hence the photomultiplier tube.}
\end{figure}

In the last few years a variety of electrostatic devices for storing and handling low energy ion beams have been designed and operated.
Electrostatic storage rings \cite{mol1997,ahnt2002,bmbt2008}, cone traps \cite{scjf2001}, \emph{Orbitraps} \cite{mak2000}, and electrostatic ion beam traps (EIBT)\cite{zhv97,ben1997}
 are now used to study atomic and molecular metastable states or molecular
 fragmentation, photodissociation  or mass spectrometry (see, e.g., Ref. \cite{ahz2004} for a review). The design and study of these instruments relies nowadays
 mainly on computer simulations. An EIBT, as designed by D.~Zajfman and collaborators \cite{zhv97,dfhr1997}, and independently by W.H.~Benner \cite{ben1997}, is a purely electrostatic
 trap composed of two electrostatic mirror--Einzel lens combinations, as represented on Fig.~\ref{trap3d}.
 This trap has many interesting features \cite{osy06}: on one hand, it offers trapping of energetic particles (\unit{}{keV})
 in a well defined direction and on the other hand it is small, relatively inexpensive and has a field-free region where ions move freely
 and where measurements can easily be performed. It can also be used as a moderate-resolution mass spectrometer \cite{sgp2002}.

In this paper, our aim is to provide a method enabling the determination
 of analytical solution to the electrostatic potential in any axially symmetric configuration using the elegant method of conformal mapping.
 We will present the method on the EIBT, but it can be used on other sets of electrodes.  

The main reason for quadrupole traps' extensive use in precision experiments, lies in the fact that their fields can easily
 be described by an analytical formula. It enables a deeper understanding of many subtle phenomena like frequency shifts due to space charge.
The inventors of the ion trap resonator used either a matrix approach \cite{sgp2002} or numerical simulations \cite{psa2002}.  The former are useful for acquiring a qualitative understanding, but 
cannot provide detailed insight in the operating conditions, while the later may not provide enough numerical accuracy to follow the particles during several tens of 
thousands of oscillations inside the trap, or be too time and resources consuming to explore many different potential configurations.
 The tracking of particles in the EIBT is rather difficult because of the combination of a long free-flight zone between
 the mirrors and of two areas in which the particles slow down, stop and reverse their course on very short distances, while being subjected to strong,  rapidly varying, electrostatic fields.
The simulation time becomes rapidly a limitation when many potential configurations must be studied while looking for new operating conditions. Instead of two tuning parameters as in a Paul trap,
 we have to fix the potentials of five electrodes in a symmetric configuration (ten when each side of the trap is set differently).
 The space to explore is therefore too large for numerical simulations in which the field is determined by usual finite element methods, where one has to make a different calculation
 for each set of parameters. Moreover numerical errors accumulate and perturb the trajectory of the particle over long trapping times.
 In the sequel, we will show how to find a formula, depending only on these parameters, which is able to give the electrostatic potential with good accuracy. We will also show some practical applications.

Before we present our method, we will just review two ways of calculating electrostatic potentials and explain why they are difficult to use in our case. 

Green's functions often yield analytical results because they allow reduction of the solution of the Dirichlet problem to the calculation of the following
 integral:
\begin{equation*}
V(M_0)=-\epsilon_0\int_S U_S(M)\frac{\partial G(M,M_0)}{\partial n}dS,
\end{equation*}
where $M_0$ is the point where the potential is evaluated, $\epsilon_0$ is the vacuum permeability, $n$ parametrizes the direction orthogonal
 to the surface, $U_S(M)$ is a given potential distribution
 over the surface S and $G$ is a Green's function. However, Green's functions are only known for simple geometries, which limits
 the analytical approach.

A very interesting method called quasi-Greens' functions has been developed in \cite{vkb08}. This method is based on the division of a complicated 
geometry into different simple shapes. The main drawback is that the final expression is given as an infinite sum whose coefficients have no closed
 form. In practice, the given expression, although analytical, is much more complicated than the one we will present in this article.
 
 Another technique, is the charge ring method \cite{cru1963, rrb1982, PRR1997, WFP2004}. The integral form of the Poisson equation is applied to $N$ rings representing the geometry. If $N$ is large enough, we get a set of linear equations $\Phi=AQ$ , where $\Phi$ is the potentials applied to the rings, $Q$ is the charge induced on each ring and $A$ is a matrix depending only on the geometry. Once the inverse of $A$ has been determined, the charge of each ring is know and the potential at a point $\bold{r}$ that is not on the boundary is given by:
\begin{equation*}
V(\bold{r})=\sum_{i=1}^{N}\frac{q_i}{4\pi \epsilon_0 s_i} \int_{s_i}\frac{d\bold{r_i}}{|\bold{r}-\bold{r_i}|}.
\end{equation*}
 where $s_i$ is the area of ring $i$. Hundreds of rings a usually provide a accuracy of the on-axis potential of the order of $10^{-4}$ (one order of magnitude better than the method proposed here, see Sec. \ref{sec:solving}). However, it is at least one order of magnitude slower to compute: even though the inverse of $A$ is calculated only once for a given geometry, one has to evaluate numerically $N$ integrals, which is much longer than to evaluate common mathematical functions.  We have implemented both methods on the same geometry and the conformal method was 64 times faster than the ring method. The choice between those two techniques will rely on the need to improve accuracy or speed.
 
This article is organized as follows: in Sec. \ref{sec:bertram}, we present an approximate method to obtain the analytical potential of an axisymmetric
 set of electrodes,
having the same radius. In Sec. \ref{sec:conformal}, we explain how to use Schwarz-Christoffel method to alter the metric and obtain a set of
 electrodes with the same radius and in
 Sec. \ref{sec:solving} we solve the problem
 in this new space using the method of section \ref{sec:bertram}. Section \ref{sec:summary} contains a summary of the
 key steps of the whole method as well as a discussion on the improvement of the accuracy. Finally, in Section \ref{sec:applications} we show that the dynamics of the ions in the EIBT is governed by an Hill's equation and we present a stability map showing what experimental parameters lead to an efficient trapping.

\section{Separation of variables and Bertram's method}
\label{sec:bertram}

We start from the Laplace equation in cylindrical coordinates:
\begin{equation}
\label{laplace}
\nabla^2V=\frac{\partial^2V}{\partial r^2}+\frac{1}{r}\frac{\partial V}{\partial r}+\frac{\partial^2V}{\partial z^2},
\end{equation}
where $V=V(r,z)$ is the potential at radius $r$ from the axis and at a distance $z$ from the center of the trap. Using the method of separation
 of variables, i.e., assuming $V(r,z)=R(r)Z(z)$, we obtain \cite{jac98}:
\begin{eqnarray}
\label{bessel}
\frac{d^2Z}{dz^2}-k^2Z&=&0 \nonumber \\ 
r^2 \frac{d^2R}{dr^2}+r \frac{dR}{dr}+r^2k^2 R&=&0,
\end{eqnarray}
where $k$ is a real constant. \eqref{bessel} is a particular case of the general Bessel equation \cite{aas1965}, whose solution is a Bessel function of first kind: $J_0(kr)$. We can use the general solution given by a Fourier-Bessel series of the form:
\begin{equation}
\label{vrz}
V(r,z)=\frac{1}{2\pi}\int_{-\infty}^{+\infty}a(k)J_0(kr)e^{ikz}dk,
\end{equation}
enabling us to take into account the boundary conditions.
Thus, if the potential at some radius $R$ is known as a function of $z$, then $a(k)$ can be found by means of the Fourier transform:
\begin{equation}
a(k)J_0(jkR)=\int_{-\infty}^{+\infty}V(R,z)e^{-ikz}dz.
\end{equation}

\begin{figure}
\includegraphics[width=0.4\textwidth]{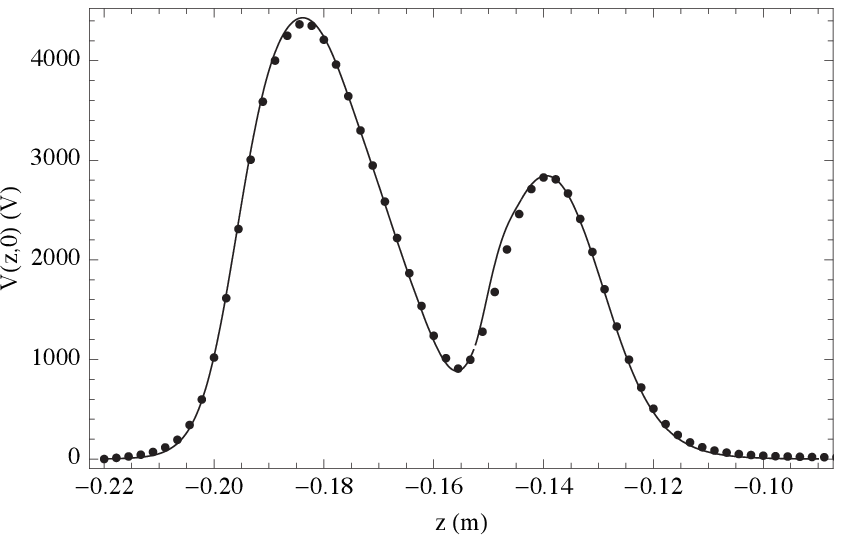}
\includegraphics[width=0.4\textwidth]{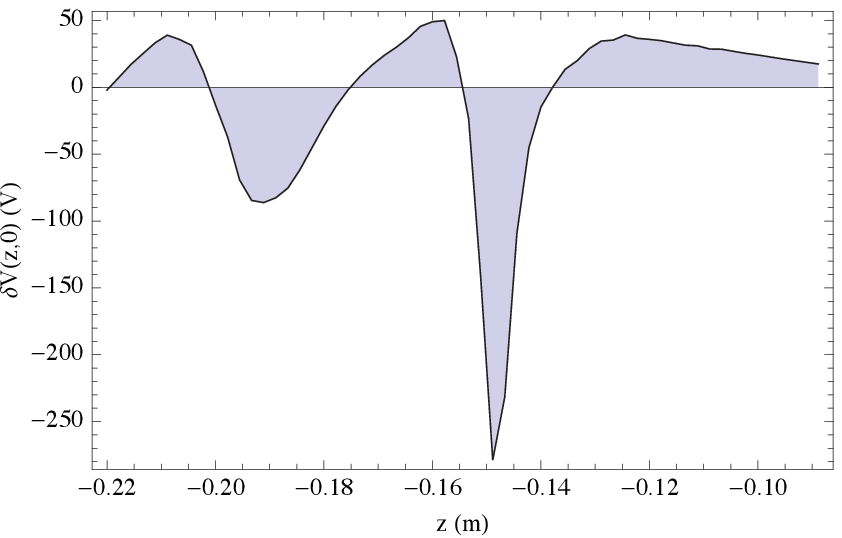}
\caption{\label{potRvar} (Color online) The line represents the analytical function and the dots is the numerical solution achieved with the finite elements software COMSOL Multiphysics® (top). The difference between the two previous curves (bottom). In this case, we applied Bertram's method replacing R by a function of z. Even though the function is a polynomial going smoothly from $R_z$ to $R$, we see that the error is large (\unit{-258}{V}) near $z=-0.15$ where the transition occurs. Here \{$V_1=4513\ V,V_2=4836\ V,V_3=3112\ V,V_4=1642\ V,V_z=,3941\ V$\}.}
\end{figure}

Given that any solution of the Laplace equation in a cylindrical symmetry is also of the general form:
\begin{equation}
V(r,z)=\sum_{n=0}^{+\infty}\frac{(-1)^n}{n!^2 2^{2n}}r^{2n}V^{(2n)}(0,z),
\end{equation}
it is sufficient to determine $V(0,z)$ along the axis.\par
Following Bertram \cite{ber40}, we assume that if we know the potential $V(R,\zeta)$ at a distance $R$ from the axis then, the potential on the axis is well approximated by the formula:
\begin{equation}
\label{bertram}
V(0,z)=\frac{\omega}{2R}\int_{-\infty}^{+\infty}V(R,z-\xi)\text{sech}^2(\frac{\omega}{R} \xi) d\xi,
\end{equation}
where the constant $\omega=4A_0=1.3152$, and $A_0$ is the first coefficient of the following development \cite{ber40}:
\begin{equation}
\frac{1}{J_0(jk)}=\sum_{n=0}^{+\infty}A_n \cos{\frac{nk}{2}}.
\end{equation}
In our case, \emph{if all the electrodes had the same radius}, we could use directly this method, taking $V(R,\zeta)$ as a piecewise linear function of the set of potentials \{$V_1,V_2,V_3,V_4,V_z$\}. However, since the radius of the first four electrodes is $R=\unit{8}{mm}$, different from the Einzel electrodes where $R_z=\unit{13}{mm}$, this method does not work in the area where the radius changes. We tried to introduce a smooth function $R(z)$ in the integral, but the difference between those results and a finite element solution always shows a large discrepancy as illustrated on Fig.~\ref{potRvar}.

The previous method is rather simple and efficient to find the potential. Its only limitation is the need to have identical radii for all electrodes.  
In the next section, we will show how to use conformal mapping to place ourselves in the space where the borders have a constant radius.

\section{Conformal mapping}
\label{sec:conformal}

Conformal mapping is widely used in applied physics and chemistry. One can cite the design of airfoils: the Joukowsky transformation \cite{tan79}
 reduces the study of the laminar flow 
on a complicated profile to the much easier study of a cylinder in the transformed flow, or the study of diffusive flow at micro-ring electrodes
 in analytical chemistry \cite{aos2004,oas2004}. Applications in electrostatic potential determination are also known (see, e.g., \cite{boe1990,rsmv1996,aos2003,we2008}). However, all these works use conformal mapping in a geometry where one dimension can be considered infinite. A section perpendicular to this infinite dimension is then mapped. In this paper, we show that \emph{conformal mapping can be used to find the electrostatic potential in a space limited axisymmetric geometry} by mapping the plane parallel to the axis and rotating it.
 
\begin{figure}
\includegraphics[width=0.3\textwidth,trim = 1.6cm 11cm 5.7cm 2cm,clip]{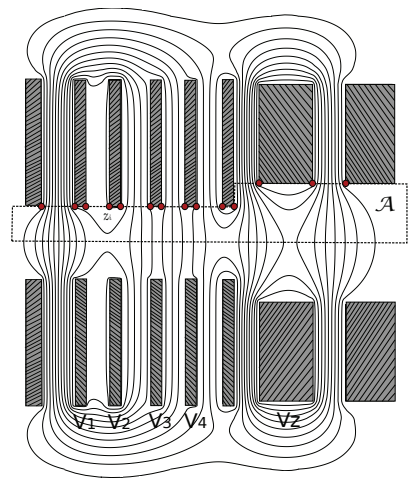}
\caption{\label{domainA} (Color online) A section of one side of the trap: the crosshatched areas are the electrodes, the continuous lines are the equipotentials, the domain $\mathcal{A}$ is surrounded by a dashed border and the circled points represent the $z_i$ points, which define our boundary conditions.}
\end{figure}

 In order to use the results of \ref{sec:bertram}, the  first step is to find the holomorphic function that maps section $\mathcal{A}$, described
 on Fig.~\ref{domainA}, on a rectangle. Holomorphic functions are of great interest because angles are conserved under those transformations: equipotential lines stay orthogonal to field lines.
 
\begin{figure}
\includegraphics[width=0.5\textwidth,trim = 8cm 14.7cm 3cm 5cm,clip]{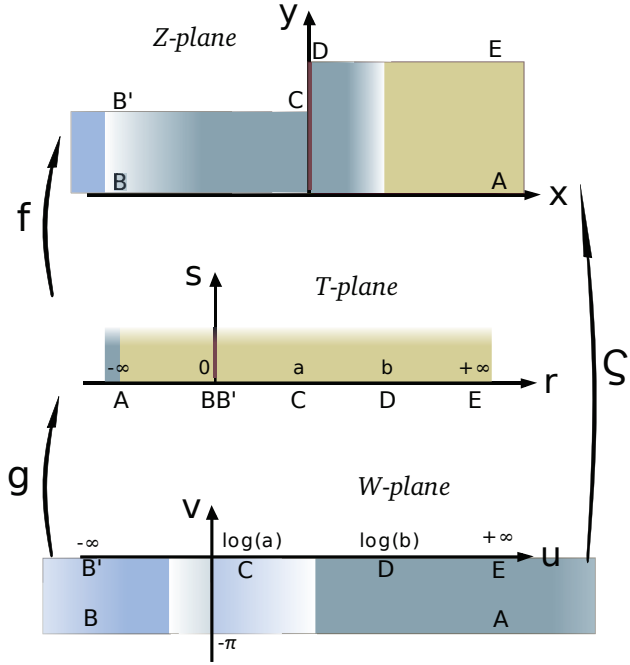}
\caption{\label{workingplanes} (Color online) The three working planes. The $Z$-plane is the physical plane corresponding to domain $\mathcal{A}$ on Fig.~\ref{domainA}.
 The $T$-plane is an intermediate step due to the fact that the Schwarz-Christoffel method always maps a polygon on the upper complex plane where $Im(t)=s>0$.
 The $W$-plane is the calculation plane where we can apply the method of section \ref{sec:bertram} since it can be seen as the section of a constant radius cylinder.
 The arrows on the border represent the three mapping used to transform one plane to the other.}
\end{figure}

We first use the Schwarz-Christoffel transformation \cite{SL03} to map domain $\mathcal{A}$, parametrized by $z=x+iy$ onto the upper
 half plane where $t=r+is$, as represented on Fig.~\ref{workingplanes}:
\begin{equation*}
\frac{dz}{dt}=f'(t)=K_1 t^{-1}(t-a)^{1/2}(t-b)^{-1/2},
\end{equation*}
so that:
\begin{equation*}
z=f(t)=K_1 \int_0^t \frac{(t'-a)^{1/2}}{t'(t'-b)^{1/2}}dt'+K_2
\end{equation*}
where $K_1$ and $K_2$ are constants to be determined. With an appropriate choice of the origin (f(0)=0), we have $K_2=0$. For $K_1$, we use the boundary conditions: going from point $A$ to point $E$ in the $Z$-plane (see Fig.\ref{workingplanes}) 
corresponds to a large semicircle of radius $\rho\to +\infty$ and $\theta$ from $0$ to $\pi$ in the $T$-plane:
\begin{equation*}
iR_z=K_1\int_{0}^{\pi}\frac{(\rho e^{i\theta}-a)^{1/2}}{\rho e^{i\theta}(\rho e^{i\theta}-b)^{1/2}}i\rho e^{i\theta}d\theta.
\end{equation*}
When taking the limit $\rho\to +\infty$, this reduces to:
\begin{equation*}
iR_z=K_1\int_{0}^{\pi}\frac{(\rho e^{i\theta})^{1/2}}{\rho e^{i\theta}(\rho e^{i\theta})^{1/2}}i\rho e^{i\theta}d\theta=K_1i\pi,
\end{equation*}
and so $K_1=R_z/\pi$.
The second boundary condition, going from $B$ to $B'$ in the $Z$-plane, is expressed by integrating around $BB'$ with $\rho\to 0$ and $\theta$ going from $0$ to $\pi$:
\begin{eqnarray*}
iR&=&\frac{R_z}{\pi}\int_{0}^{\pi}\frac{(\rho e^{i\theta}-a)^{1/2}}{\rho e^{i\theta}(\rho e^{i\theta}-b)^{1/2}}i\rho e^{i\theta}d\theta\\\\
&=&\frac{R_z}{\pi}\sqrt{\frac{a}{b}}\int_{0}^{\pi}id\theta\\\\
&=&iR_z\sqrt{\frac{a}{b}}
\end{eqnarray*}

Choosing $a=1$ only fixes the origin in the $T$-plane, and it implies $\sqrt{b}=\frac{R_z}{R}$. Finally, we make the substitution:
\begin{equation*}
p=\sqrt{\frac{t'-1}{t'+1}}
\end{equation*}
and the integration gives:
\begin{equation*}
z=\frac{R_z}{\pi}\left(\frac{1}{\sqrt{b}}\ln{\frac{\sqrt{bp}-1}{\sqrt{bp}+1}}+\ln{\frac{1+p}{1-p}}\right).
\end{equation*}
The mapping from the W-plane to the T-plane is much simpler:
\begin{equation*}
w=g(t)=Log(t)
\end{equation*}
and we can finally link the $Z$-plane to the W-plane by the following transformation:
\begin{eqnarray}
\label{zeta}
z&=&f(g^{-1}(w))\nonumber \\
&=&\frac{R_z}{\pi } \left(\log
   \left(\frac{\sqrt{e^w-b}+\sqrt{e^w-1}}{\sqrt{e^w-b}-\sqrt{e^w-1}}\right)\right. \nonumber \\
 && \left. +\frac{1}{\sqrt{b}}\log \left(\frac{\sqrt{b
   \left(e^w-1\right)}-\sqrt{e^w-b}}{\sqrt{b
   \left(e^w-1\right)}+\sqrt{e^w-b}}\right)\right)\\
   &=&\zeta(w)\nonumber,
\end{eqnarray}
The transformation $z=\zeta(w)$ gives a one-to-one mapping between points of the $Z$-plane and points of the $W$-plane and it is a conformal transform as the composition of two conformal transforms. Since $\zeta$ is not invertible, we have found a good approximating function defined on three intervals and inverse of which is:
\begin{equation*}
\begin{array}{l}
\zeta^{-1}(z)=  \\\\
\left\{
\begin{array}{lr}
-\log{\frac{(1-b)(\exp{[-\sqrt{b}(\frac{\pi z}{R_z}-\log{\frac{\sqrt{b}+1}{\sqrt{b}-1}}})]+1+b)}{4b}}&z<-z_0\\\\
P(z)&-z_0\le z\le z_0,\\\\
\log{\frac{1-b}{4}(\exp{[\frac{\pi z}{R_z}-\frac{1}{\sqrt{b}}\log{\frac{\sqrt{b}-1}{\sqrt{b}+1}}]}+\frac{1+b}{1-b})}&z_0<z,
\end{array}\right.
\end{array}
\end{equation*}
where $z_0=\unit{0.007}{mm}$ and $P(z)$ is a 6th-order polynomial connecting continuously the two asymptotic expansions. The points at $\pm z_0$ represent the limit of validity of the two asymptotic expansions. 

\begin{figure}
\includegraphics[width=0.5\textwidth]{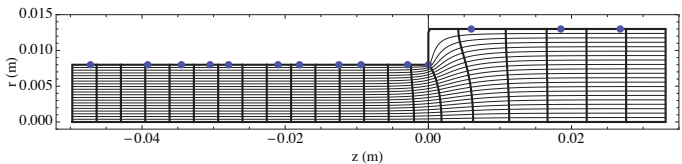}
\caption{\label{mapping} (Color online) This figure represents the image of the mapping $\zeta(w)$ applied to the rectangle defined as $w=u+iv$ with $-20<u<8$ and $-\pi<v<0$. The dots indicate the position of the $z_i$. }
\end{figure}

There is one last thing to do before one can solve the problem in the $W$-plane. Fig.~\ref{mapping} shows the image of the rectangle defined as $w=u+iv$ with $-20<u<8$ and $-\pi<v<0$. On the vertical bold lines, $u$ is constant and we will call these lines iso-$u$. We see that the iso-$u$ are distorted near $0$. Therefore, the distance between the points $w_i=\zeta^{-1}(z_i)$ is not the same as the distance between the points $z_i$: the metric is not conserved on the border. Consequently, the width of the electrodes near the point $C$ in the $W$-plane is not the physical width, which leads to errors in the potential. In order not to modify the metric on the border, one should first apply the function
\begin{equation}
\beta(z)=\zeta(\zeta^{-1}(z)-i\pi)
\end{equation}
to the points $z_i$ so that their image have the same distance between them in the $W$-plane and in the $Z$-plane (physical plane). This function uses the fact that, on the axis, the metric is not modified from one plane to the other. It is the same idea as in \cite{aos2003} where a "space dependent diffusion coefficient" is used to account for the fact that real space is compressed/expanded unevenly to fit the $W$-plane. In the sequel, we shall use the following notation: $\tilde{z_i}=\beta(z_i)$. 
\subsection{Solving in the W-plane}
\label{sec:solving}
Using $\zeta(w)$ we can now apply Bertram's method in the W-plane where the radius is constant.
Following \cite{ber40}, we assume that the potential varies linearly between the electrodes. We shall come back to this approximation in the last part of this section. We now use \eqref{bertram} in the $W$-plane where the radius is constant $R=\pi$. On the border, we have:
\begin{equation*}
V_W(\pi,u)=\sum_{i=1}^{13}\left(\frac{V_{i+1}-V_i}{u_{i+1}-u_i}(u-u_i)+V_i\right)\Pi(u_i\to u_{i+1}),
\end{equation*}
with $u_i=\zeta^{-1}(\tilde{z_i})$, where $\tilde{z_i}=\beta(z_i)$. $z_i$ are the positions of the points on Fig.~\ref{domainA}, $V_i$ is the potential at $z_i$. $\Pi(u_i\to u_{i+1})$ is a function equal to zero everywhere except between $u_i$ and $u_{i+1}$ where it is equal to one. Replacing in \eqref{bertram}, we obtain:
\begin{eqnarray}
\label{bigone}
V_W(0,u)&=&-\frac{1}{2}\sum_{i=1}^{13}\left[\frac{}{}Q_i(\phi_{i+1}(u)-\phi_i(u))\right.\nonumber\\
&&+(Q_i u_i-V_i)\chi_i(u) \\
&&\left.+\frac{\pi Q_i}{\omega}\psi_i(u)\right]\nonumber,
\end{eqnarray}
where:
\begin{eqnarray}
\label{bigone-elem}
Q_i&=&\frac{V_{i+1}-V_i}{u_{i+1}-u_i}\nonumber\\
\phi_i(u)&=&u_i \text{tanh}\frac{\omega}{\pi}(u-u_i)\nonumber\\
\chi_i(u)&=&\frac{2\text{sinh}\frac{\omega}{\pi} (u_{i+1}-u_{i})}{\text{cosh}\frac{\omega}{\pi}(2u-u_{i+1}-u_i)+\text{cosh}\frac{\omega}{\pi}(u_{i+1}-u_i)}\nonumber\\
\psi_i(u)&=&\text{log}\left(\frac{\text{cosh}\frac{\omega}{\pi}(u-u_{i+1})}{\text{cosh}\frac{\omega}{\pi}(u-u_i)}\right)\nonumber.
\end{eqnarray}

Now that we have $V_W$, the potential along the axis in the $Z$-plane is given by:
\begin{equation*}
V_Z(0,z)=V_W(\zeta^{-1}(z))
\end{equation*}
as $w=\zeta^{-1}(z)$ varies from point $B$ to point $A$. 

\begin{figure}
\includegraphics[width=0.4\textwidth]{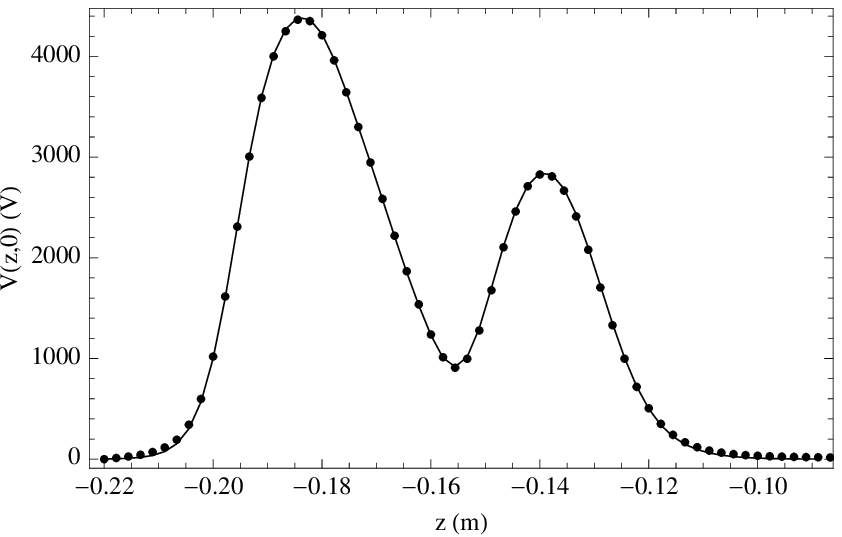}
\includegraphics[width=0.4\textwidth]{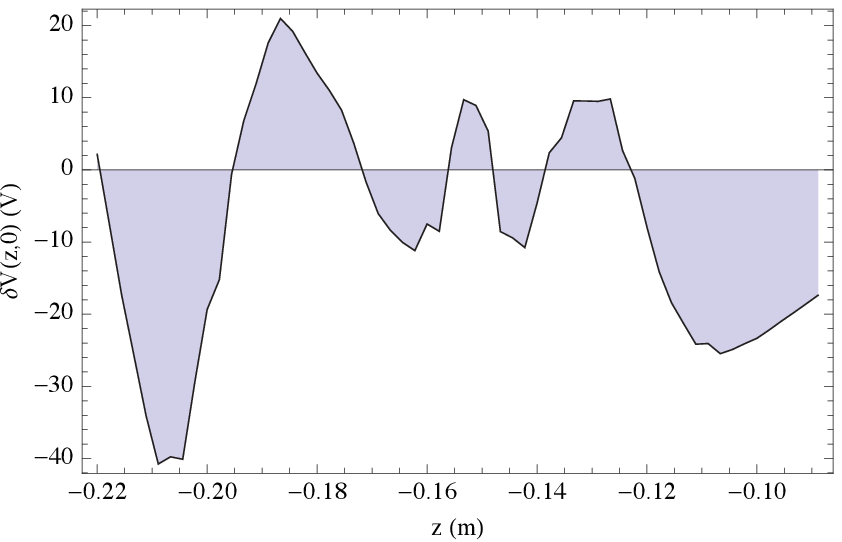}
\caption{\label{potConf} (Color online) The line represents the analytical function and the dots is the numerical solution (top). The difference between the two previous curves (bottom). We see that the error is around 1\% and its smooth repartition shows that it does not arise from the varying radius. This difference is the consequence of the approximation we used to fix $V(R,z)$ between the electrodes. Here \{$V_1=4513\ V,V_2=4836\ V,V_3=3112\ V,V_4=1642\ V,V_z=,3941\ V$\}.}
\end{figure}

\begin{figure}
\includegraphics[width=0.4\textwidth]{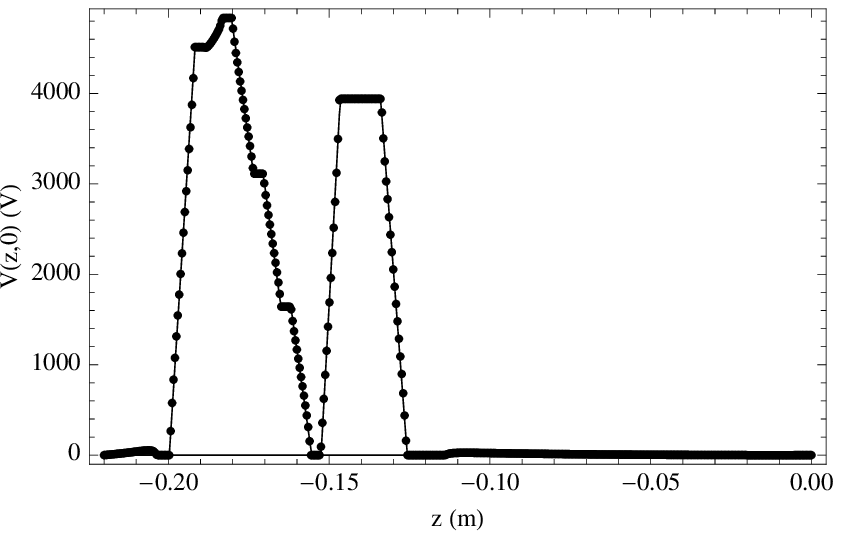}
\includegraphics[width=0.4\textwidth]{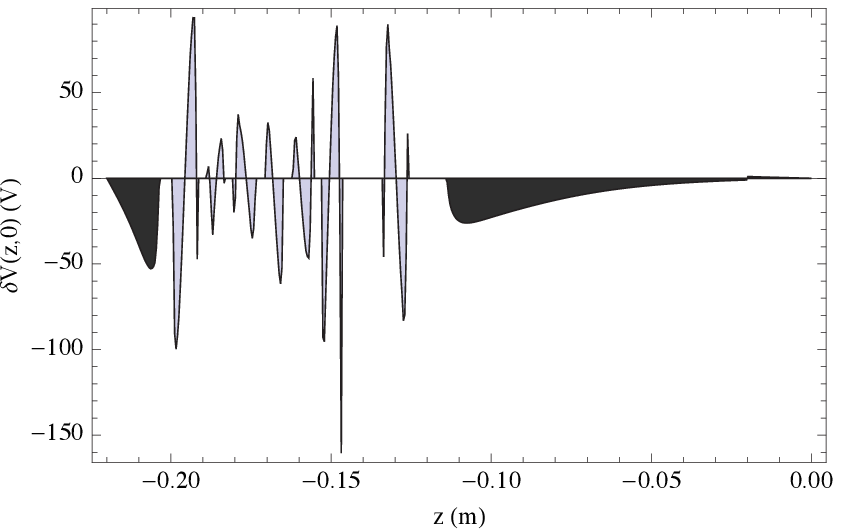}
\caption{\label{potBorder} The line represents the potential on the border used to calculate the potential on the axis and the dots is the numerical solution achieved with Comsol multiphysics (top). The difference between the two previous curves (bottom). Here \{$V_1=4513\ V,V_2=4836\ V,V_3=3112\ V,V_4=1642\ V,V_z=,3941\ V$\}.}
\end{figure}

The error between the numerical solution and our result is around 1\%, which is enough for many applications. There are two main sources of errors. First, the calculation does not take into account the field leaking at the front and at the rear of the set of electrodes. It can be seen on Fig.~\ref{domainA} that the potential is not exactly zero on the axis after the neutral electrodes preceding $V_z$ and following $V_1$. This explains the two dark negative zones on Fig.~\ref{potBorder}. Second, on the same figure, we see that the hypotheses we have made, concerning the variation of the potential between the electrodes, induce an error of about \unit{100}{V} on the border. Yet, since this error is oscillating along each border, it compensates and the error on the axis is only about \unit{20}{V} (Fig.~\ref{potConf}). An attempt to enhance the potential at the borders is described in \cite{ber42}. Finally, there is an error coming from the cylindrical term of Eq. (\ref{laplace}), which is not invariant under conformal mapping. However, it is always at least one order of magnitude smaller than the two previous error types.

\begin{figure}
\includegraphics[width=0.5\textwidth,trim = 2cm 1cm 2cm 2cm,clip]{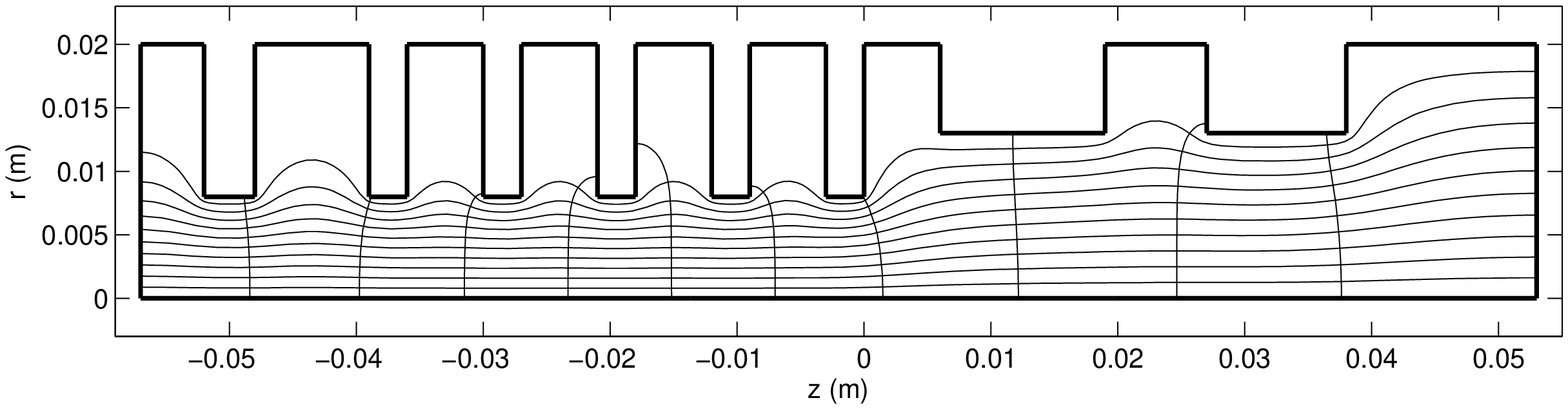}
\caption{\label{mapping2} This figure represents the image of a mapping obtained numerically.}
\end{figure}

The best way to increase accuracy is to use a domain that takes into account the shape of the electrodes. Instead of the mapping of Fig.~\ref{mapping}, we could have used the one of Fig.~\ref{mapping2}. Besides showing that this method can be applied to complicated geometries, Fig.~\ref{mapping2} enables us to emphasize a crucial point: the mapping might not be analytic for a given geometry. However, since the geometry does not change, it is enough to calculate the inverse map numerically one time, find the polynomial that fits this numerical solution on the axis and then, using Bertram's formula we have an analytic expression whose parameters are the electrodes' potentials. With this mapping, the accuracy is approximately 0.1\%. 

Solving numerically a Schwartz-Christoffel map will not be treated here because we chose a complete analytical case to show all the details of the method. One could refer to \cite{SL03, DRV1998} for an extensive review on all the numerical techniques involved. Note that all the results presented in the sequel, use the simple case of Fig.~\ref{mapping2}.

\subsection{Summary of the method}
\label{sec:summary}
\begin{enumerate}
\item Write the the Schwartz-Christoffel transformation adapted to the geometry to obtain $z=\zeta(w)$. For more details see \cite{SL03}. \\
\item If the inverse $w=\zeta^{-1}(z)$ is not straightforward, it can be well approximated by a polynomial.\\
\item Let $z_i=x_i+iR(x_i)$ be the points of the $Z$-plane defining the position of each side of the electrodes, one has to apply the function $\beta(z)=\zeta(\zeta^{-1}(z)-i\pi)$ to obtain the set of points $\tilde{z_i}=\zeta(\zeta^{-1}(z_i)-i\pi)$\\
\item Transpose the problem from the $Z$-plane to the $W$-plane where $w_i=\zeta(\tilde{z_i})=u_i+iv_i$.\\
\item Use the formula (\ref{bigone}) to obtain $V_W(u,-Pi)$.\\
\item The potential of one point $z=x+0\times i$ on the axis in the $Z$-plane is $V_Z(z,0)=V_W(\zeta^{-1}(z))$.
\end{enumerate}

\section{Application: stability map}
\label{sec:applications}
The main difficulty to tune an EIBT is due to the large number of parameters implied in its manipulation: five potential values on each side, the energy of the ions, their temperature and their charge-to-mass ratio. The designers of the trap used an optical model consisting of mirrors and lenses \cite{osy06}, yet, since the focal length is not linked to the values of the potential, the behavior of the EIBT, given a set of potentials, is unpredictable. Until now, experimentalists had to use simulation software like SIMION® \cite{cpo} to determine optimal trapping conditions \cite{osy06}. Since a finite element calculation has to be achieved each time a parameter is changed, this method turns out to be fastidious. Worse, simulating the trajectory of ions going back and forth is much more difficult than simulating a beam because the errors on the position accumulate and become comparable to the size of the trap. Even with the smallest step and a recursive method, SIMION® \cite{cpo} was not able to enforce energy conservation after a few hundreds oscillations (the error was about \unit{100}{eV}).\\
\begin{figure}
\includegraphics[width=0.5\textwidth]{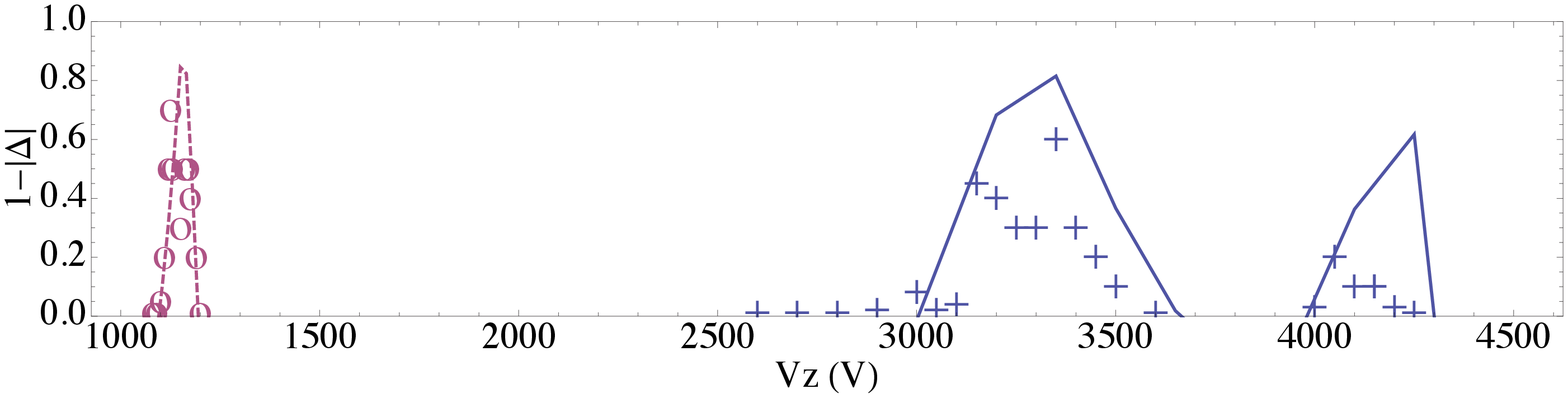}
\caption{\label{stab_vz} (Color online) These curves show $1-|\Delta|$ as a function of $V_z$ (potential applied to the lens) in two distinct cases: (right curve, +) is the trapping efficiency of $Ar^+$ at \unit{4.2}{keV}  with the same conditions as in \cite{psh2002}. The set of points is an reproduction of the experimental data in their Fig.3 a. (left curve, o) represents the trapping efficiency of $Ar^+$ at \unit{1.2}{keV} with the same conditions as in \cite{osy06}. The set of points is an reproduction of the experimental data in their Fig. 8. There is very good agreement between theory and experiment.}
\end{figure}
From Eq. (\ref{vrz}), we can limit ourselves to the second order. Higher order terms can be neglected as long as $r<\unit{8}{mm}$, which is the aperture of our trap. The potential in the trap is then given by 
\begin{equation}
\Phi(r,z)=V(z)-\frac{1}{4}r^2\frac{d^2V(z)}{dz^2},
\end{equation}
where $V(z)$ is the potential along the axis. It follows that the trajectory of one ion in the trap is determined by the following set of equations:
\begin{eqnarray}
\label{pfd}
\frac{m}{q}\frac{d^2z}{dt^2}=-\frac{dV(z)}{dz}+\frac{1}{4}r^2 \frac{d^3V(z)}{dz^3}\\
\label{pfd2}
\frac{m}{q}\frac{d^2r}{dt^2}=\frac{1}{2}r \frac{d^2V(z)}{dz^2}
\end{eqnarray}
The second term of the right side of equation (\ref{pfd}) is small compared to the first (for $r<8$ mm) and can thus be neglected. We obtain the longitudinal motion of the ion $z(t)$. Substituting $z(t)$ in (\ref{pfd2}), we obtain an Hill's equation \cite{hill1886}:
\begin{equation}
\label{hill}
\frac{d^2r}{dt^2}-\left(\frac{q}{2m} \frac{d^2V(z)}{dz^2}\vline_{z(t)}\right) r=0,
\end{equation}
the term in parentheses being a periodic function of period $T$.
This equation arose in the study of the moon's dynamic and the usual method to discuss the stability of its solution consists of calculating infinite determinants \cite{whi14}. The principal matrix of this equation is:
\begin{equation}
M(t)=\left( 
\begin{array}{cc}
\psi_1(t,t_0)&\psi_2(t,t_0)\\
\dot{\psi_1}(t,t_0)&\dot{\psi_2}(t,t_0)
\end{array}
\right)
\end{equation}
where $\psi_1(t,t_0)$ is the solution of (\ref{hill}) with initial conditions $\psi_1(t_0,t_0)=1$ and $\dot{\psi_1}(t_0,t_0)=0$ and $\psi_2(t,t_0)$ with $\psi_2(t_0,t_0)=0$ and $\dot{\psi_2}(t_0,t_0)=1$. Liouville's formula \cite{tes08} shows that:
\begin{equation}
\det M(t,t_0)=1
\end{equation}
and therefore the characteristic equation of the monodromy matrix $M(t_0+T)$ is given by \cite{tes08}:
\begin{equation}
x^2-2\Delta x+1=0
\end{equation}
where:
\begin{equation}
\Delta=Tr(M(t_0+T))=\frac{\psi_1(t_0+T,t_0)+\dot{\psi_2}(t_0+T,t_0)}{2}
\end{equation}
Applying Floquet's theorem \cite{mw1979}, we know that if $\Delta^2>1$, one of the two solutions is unbound but if $\Delta^2<1$ there are two solutions:
\begin{equation}
r(t)=e^{\pm\gamma t}p_{\pm}(t)
\end{equation}
where $p_{\pm}(t+T)=p_{\pm}(t)$, $ \gamma=Im(\frac{1}{T} \log(x_+))$ and $x_+=\Delta + \sqrt{\Delta^2-1}$. In conclusion, Hill's equation is stable, and thus trapping can be observed, when $| \Delta |<1$. \\
We now compare these theoretical results with experiment. 

Figure~\ref{stab_vz} shows a comparison with the results published by other groups using the same kind of trap. These curves show $1-|\Delta|$ as a function of $V_z$ (potential applied to the lens).They show a perfect agreement with two independent groups \cite{psh2002}, \cite{osy06}. We also tried to reproduce the data of \cite{ACK2009} using $H_2^+$ at \unit{1.0}{keV} and a negative potential on the Einzel lens. We also predict three stability intervals, approximately at the same position, because we have not taken into account the differences in the geometry of their trap. However we notice that the method works also with negative potentials. Our method enables to plot these curves in less than a second whereas the cited authors had to make a SIMION® \cite{cpo} simulation for each point.\\

\begin{figure}
\includegraphics[width=0.4\textwidth]{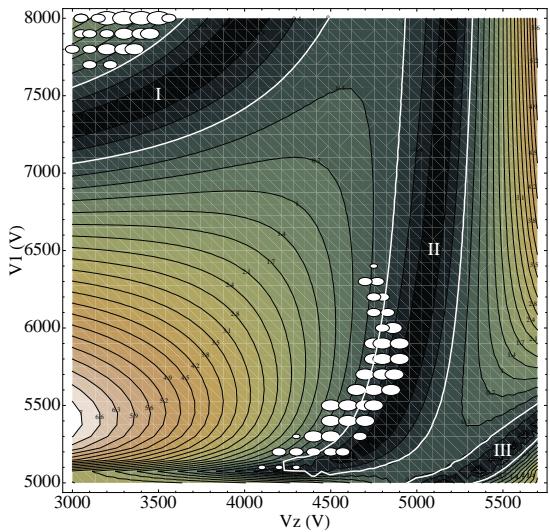}
\caption{\label{v1vz} (Color online) The contours show the iso-$\eta$ values where $\eta=|\Delta|-1$. The stability region is defined by $\eta<0$. The white dots indicate settings where trapping was experimentally observed. The radius of each dot is proportional to the trapping efficiency. The thick white line is the border of the three stability region: there are three stability zones marked I, II and III corresponding to the three types of orbits of Fig.~\ref{orbit}. We used $O^{4+}$ at \unit{5.2}{keV/charge} and the potential set is: \{$V_1,V_2=5850\ V,V_3=4150\ V,V_4=1650\ V,V_z$\}.}
\end{figure}

\begin{figure}
\includegraphics[width=0.3\textwidth]{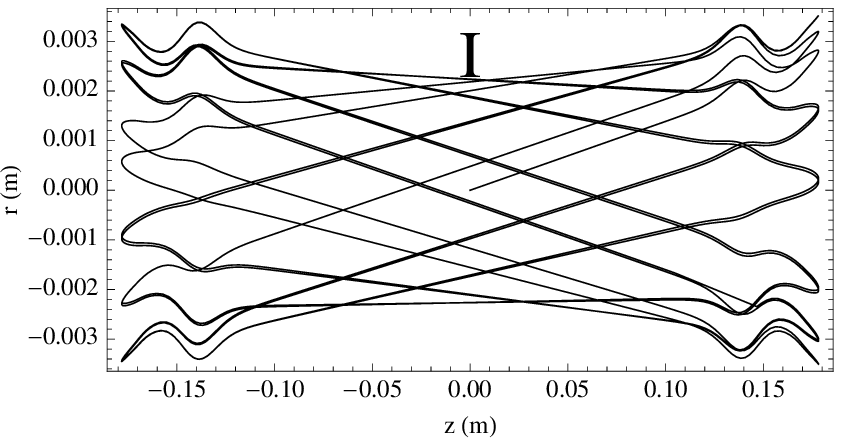}
\includegraphics[width=0.3\textwidth]{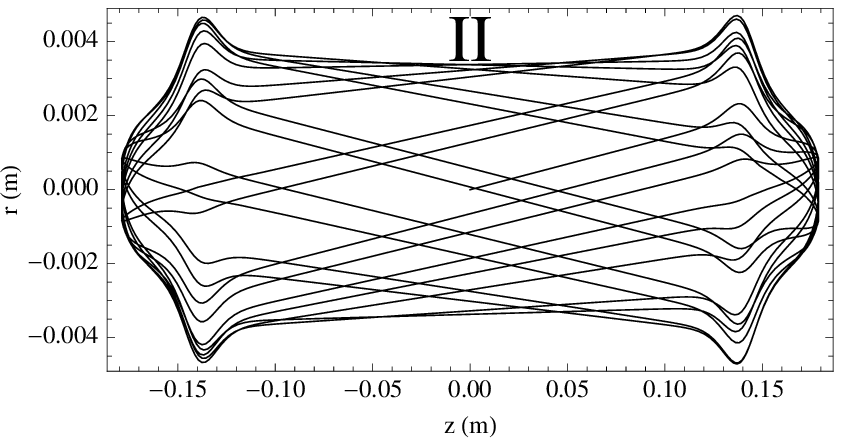}
\includegraphics[width=0.3\textwidth]{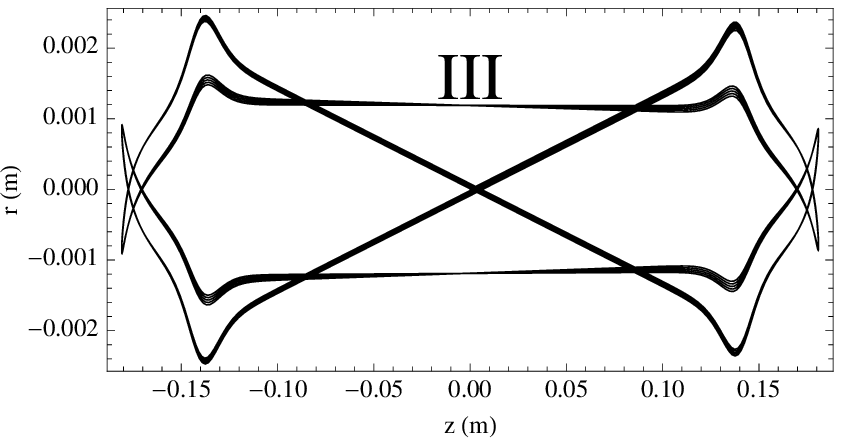}
\caption{\label{orbit} Trajectories of $O^{4+}$ at $5.2\ kV$ (top): Stability zone I: \{$V_1=7500\ V,V_2=5850\ V,V_3=4150\ V,V_4=1650\ V,V_z=3650\ V$\} . (middle): Stability zone II: \{$V_1=6500\ V,V_2=5850\ V,V_3=4150\ V,V_4=1650\ V,V_z=4950\ V$\}. (bottom): Stability zone III: \{$V_1=5100\ V,V_2=5850\ V,V_3=4150\ V,V_4=1650\ V,V_z=5400\ V$\}.}
\end{figure}

Figure~\ref{v1vz} shows the stability map depending on two parameters $V_1$ and $V_z$ (the respective potentials of the rear electrode and of the Einzel lens). This is the equivalent to the famous Ince-Strutt diagram used to tune quadrupolar traps \cite{MGW05}. 
The white dots indicate settings where trapping was experimentally observed. We fixed the value of $V_1$ and scanned $V_z$. Trapped ions go through a ring at the center of the trap and induce a current. We then analyze this amplified current with a spectrum analyzer: if we see a peak corresponding to the oscillating movement of the ions, we mark this position with a white dot. The radius of each dot is proportional to the trapping efficiency. \\
There is a shift between theory and experiment, which was first thought to be caused by a technical problem: it is difficult to monitor with a good accuracy our power supplies up to \unit{8}{kV}, especially because they raise in a few nanoseconds. However, a recent improvement of the model, taking space charge effects into account, seems to explain this shift. These results will be presented somewhere else. Nonetheless, we could not explain the absence of trapping on the upper part of the stability zone II. We tested to see if higher order terms of Eq. (\ref{vrz}) could account for this, without success. No trapping can be seen in zone III, but as shown in Fig.~\ref{orbit}, this region corresponds to very peculiar closed trajectories, which seem more theoretical than observable.

\section{Conclusion}
In this article, we have given a method to calculate analytical solution to the Laplace equation in axially
 symmetric devices. This method is very general and can be used for various parts constituting a beam line.
 We successfully applied it to the EIBT and showed that the electrostatic potential is given by a formula depending on the five electrodes potentials.
 The formula has been compared with many finite element calculation where the potentials have been changed on the whole achievable range (from $\unit{0}{V}$ to $\unit{8000}{V}$ for each of the five potentials) and, in the region where the ions can move, the error is never greater than 1\%. This analytical expression of the potential inside this trap provides users with a much more powerful tool to study and optimize this novel kind of trap. As an example, we study the stability of the trap and show that they agree with experiments giving a fast and easy way to predict trapping parameters.

\section*{Acknowledgments}
Laboratoire Kastler Brossel ``Unité Mixte de Recherche du CNRS, de l'École Normale Supérieure et de l'Université Pierre et Marie Curie No.
8552''. This work was  partially supported by the Helmholtz Alliance contract HA216/EMMI and a grant from ``Agence Nationale pour la Recherche (ANR)'' number
\emph{ANR-06-BLAN-0223}. This work is performed in the framework of the SPARC collaboration 
\url{http://www.gsi.de/fair/experiments/sparc/index_e.html}. We also thank Dina Attia and Scilla Szabo for the installation of the trap and the detection scheme.

\bibliography{ref}

\end{document}